\documentstyle[epsfig,prl,aps]{revtex}

\begin{document}

\large\normalsize

\title{Massive relic neutrinos in the galactic halo and the knee in the
cosmic ray spectrum}
\author{M.T.Dova and L.N.Epele}
\address{Departamento de F\'{\i}sica, Universidad Nacional de La Plata,
C.C.67 - 1900 La Plata, Argentina}
\author{J.D.Swain}
\address{Department of Physics, Northeastern University, Boston, MA
02115}

\maketitle

\begin{abstract}

Despite many efforts to find a reasonable explanation, the origin 
of the "knee" in the cosmic ray spectrum at  $E \approx 10^{15.5} eV$ remains 
mysterious. In this letter we suggest that the "knee" may be due 
to a GZK-like effect of cosmic rays interacting with massive neutrinos 
in the galactic halo. Simple kinematics connects the location of the 
"knee" with the mass of the neutrinos, and, while the required 
interaction cross section is larger than that predicted by the 
Standard Model, it can be accommodated by a small neutrino magnetic 
dipole moment. The values for the neutrino parameters obtained from 
the analysis of existing experimental data are compatible with 
present laboratory bounds.

\end{abstract}


The cosmic ray  (CR) spectrum is well-described by a  power law of the
$E^{-\gamma}$ form with a spectral index , which is nearly constant 
over rather wide
ranges of energy  $E$,  but displays significant discontinuities. The most
notable occurs  in the  region of the  so-called "knee" 
\cite{ku,na,da,ea,am,ch,gl},  
at about  $10^{15.5}$ eV, where  the spectral index changes from  
2.75 to 3.  

Several
models have  been proposed to explain  this "knee", none  of which has
been particularly convincing. They include a change in CR sources from
Type  I  to Type  II  supernovae and/or  to  changes  of the  particle
acceleration efficiency  as a function  of the electric charge  of the
primary \cite{la,dr,pe}. Other  proposals associate changes of  the spectral index
with a reduction  in the efficiency of the  galactic magnetic field to
confine  the  CR's \cite{wd,pt},  allowing  those  at higher  
energy  to  leak
away. None of  these models is able to reproduce  the sharpness of the
"knee". Another, rather controversial, explanation of this kink in the
spectrum is to  attribute it to a recent,  strong single source\cite{tuall}.  

In
this  Letter, we  propose  that the  "knee"  may be  due to  inelastic
collisions between primary cosmic  ray protons and the cosmic neutrino
background. For light Standard Model neutrinos with mass ($m_\nu < 1 $MeV ), 
Big Bang
cosmology  predicts a  thermal neutrino  background of  temperature $T_\nu = 1.6 \times 10^{-4}$eV
with  an  average   number  density  of $n_\nu \approx 337 {\mathrm{cm}}^{-1}$
,  but   which  can  rise
dramatically in galaxies due to gravitational clustering provided that
the neutrinos  are sufficiently massive  (non-relativistic).  

The idea
is a simple extension of  the GZK mechanism \cite{gzk1,gzk2} 
whereby protons above
about  $10^{19}$eV rapidly lose energy via inelastic collisions with cosmic
background radiation photons, but  now considering the cosmic neutrino
background instead.  Even allowing for the  possibility that neutrinos
have  mass and  cluster significantly,  it is  easy to  show  that the
Standard Model weak interaction cross sections are too small for these
neutrinos   to   explain   any   structure  as   pronounced   as   the
"knee".  Fortunately,  allowing  for  a  nonzero  neutrino  mass  also
immediately  opens  up  the  possibility that  neutrinos  might  carry
magnetic  dipole  moments   which  could  significantly  increase  the
inelastic  proton neutrino  cross  section. Effectively  we propose  a
softer  GZK-like effect  where interaction  with the  cosmic microwave
background radiation  is replaced by interaction  with virtual photons
coupling to the magnetic  dipole moment of massive neutrinos. Combined
with  the increase  in  relic neutrino  density  due to  gravitational
clustering, even  rather modest values of magnetic  dipole moment turn
out to  fit the experimental  data of the KASCADE  experiment\cite{kas1}
 rather
well.  

The required neutrino mass is set by simple kinematics from the
requirement that it be sufficiently  large that the process  
$\nu \, + \, p \rightarrow \nu \,+ \Delta,\,\,\,\,\,\, \Delta \rightarrow p
\,+\, \pi$, can take
place.  Including higher  resonances  is not  expected  to change  the
results  significantly\cite{ours}.    

In  order  to   
calculate  the  inelastic
proton-neutrino cross  section in  the relevant kinematics  regime, we
use    the   parameterization    of    the   measured    quasi-elastic
(i.e. including, and dominated by the $\Delta$) cross section from \cite{Mo}
and adapt
it to the case of a neutrino with a magnetic moment. This necessitates
replacing  the  usual electromagnetic  coupling  $-ie\gamma^\mu$ 
with the  appropriate
derivative   coupling $\frac{\kappa}{2m_\nu}\sigma_{\mu\nu}q^\nu$ where
$\kappa $ is the   magnetic   moment.   A  reasonably
straightforward  calculation   described  in  more   detail  elsewhere\cite{Longerpaper} then  gives the required
cross section as a function of  both  $m_\nu$ and  $\kappa$. 
A (CP-violating) electric
dipole moment  would have the same  effect, and would  be modeled with
the  same sort  of term  with an  additional factor  of  $\gamma_5$ .   

Fixing the
threshold   energy  for  the   pion  production   at $E_{p} = 3\times 10^{15} $eV,   
one  finds $m_{\nu} \approx 100$
eV.  Laboratory experiments  clearly  rule  out such  a  mass for  the
electron  neutrino, although  the  muon and  tau  neutrinos are  still
viable candidates,  as would be some  other hypothetical neutrino-like
dark matter  particle. While neutrino  oscillation experiments suggest
the possibility of a lighter neutrino, and cosmological arguments also
tend  to favour lighter  neutrinos, this  rather large  mass is  by no
means  excluded. Of  related interest  is the  "Z-burst  model", which
seeks to explain  the origin of CR events with  energies above the GZK
cutoff  by  interaction of  ultrahigh  energy  neutrinos with  massive
neutrinos in the galactic halo,  producing bosons which then decay and
give rise to  the particles which we observe on  earth (see the recent
review \cite{rei}  and references  therein). This  model favours  a  much lower
neutrino  mass, but  both the  Z-burst  model and  the explanation  we
propose  here  can  peacefully  coexist, assuming  different  neutrino
species (of different masses) are responsible for the two effects. 

 To
estimate  the effects  of gravitational  clustering, we  use  the well
accepted galactic  halo mass distribution model from \cite{halo} where the halo
is  described by the  spheroidal density  distribution.
This  assumes a
uniform distribution  of neutrinos  in a  core of 10  kpc, and for 
$m_{\nu}=100$ eV
yields  $n_{\nu} = 1.4 \times 10^8 {\mathrm{cm}}^{-3}$,  compatible  
with  bounds  due to  the  Pauli  exclusion
principle  and the  Tremaine-Gunn  phase-space density  constraints \cite{tr}.

Following earlier work \cite{ours} on energy loss of cosmic  rays in the cosmic
microwave background, we find an energy loss rate of
\begin{equation}
dE/dt = 
\frac{c}{\gamma} \int_0^{w_{m}} dw_o\, K(w_o) \, \sigma(w_o) \, n
\end{equation}
where $w$ stands for  the neutrino final energy in  the proton rest frame,
$n$ is neutrino density, $\sigma $   the proton-neutrino cross section 
calculated above,  and  $K$ is  the  average energy  
loss  of the  nucleon in  the
collision.  

This equation is solved numerically, with the neutrino
mass,   magnetic   dipole   moment   and   propagation   time   as
parameters. Conservation of the  number of nucleons is enforced by
a balance equation
\begin{equation}
\frac{\partial N}{\partial t} = \frac{\partial [b \, N]}
{\partial E} + D \,\, \nabla^2 N + Q 
\end{equation}
Here $b(E)$  is  the mean rate at  which particles lose  energy. The diffusion
effect  due  to galactic  magnetic  field,  modeled  by the  
$D \,\, \nabla^2 N $   is
approximated  here  from the  galactic  residence  time for  CR's,
calculated from a diffusion  model with a containment volume which
extends out 10 kpc  from the core. The range  of values we consider
here is limited  from above by the age of  the galaxy ($10^{10}$ years)
and  from below  by $10^8$  years  from the  expected residence  time
derived  from spallation \cite{lon}.   The  third term  corresponds to  the
particle  injection rate  which is  assumed  to have  a power  law
behavior, so that  $Q= K \,E^{-\gamma }$.  
The solution of this equation can be obtained
in the  same manner as in  reference \cite{ours}. 

 Figure 1  shows the total
cosmic ray differential flux  as measured by KASCADE together with
the modified total energy spectrum which is obtained from a sum of
a proton  component (abundance $\approx$ 60\%)  plus an iron  component, both
with spectral index $\gamma = 2.8$ and a curve for the fitted value of 
$\kappa = (5.4 \pm 0.6)\times 10^{-6} \mu_B$, where $\mu_B$ is
the Bohr  magneton assuming a  residence time of  $3 \times 10^8$ years. 
Different
values of $\kappa$ and  $t$ fit equally  well provided that $\kappa^2t$ is 
held constant. We
use  a simple superposition  model to  estimate the  energy losses
suffered by iron nuclei due  to interactions with neutrinos in the
halo  so that  any  effect in  the  iron spectrum  should also  be
present  at higher  energies  scaling with A .  

Interestingly,  the
corresponding  soft GZK-like  cutoff  for the  heavy component  is
expected  above  $10^{17}$eV,  consistent  with  a  possible  "second
knee"\cite{yos}. The KASCADE collaboration  has shown\cite{kas2} 
that the "knee" is
dominated  by the light  component ($\approx$ 70 \%) of the  CR with  
an energy
dependent mass composition favouring  a decrease of light elements
above the  "knee". They have also  shown that the  light and heavy
mass groups  have comparable slopes  up to the "knee"  region, but
beyond this  energy the light  component follows a  steeper curve.

The  heavy mass  composition shows  no significant  "knee"  with a
constant index  $\gamma$ . The calculated  spectrum successfully reproduces
the  KASCADE  data  in  the  region  of  the  sharp  "knee",  with
abundances   matching   well   with   those   estimated   by   the
collaboration. The differences  between the predicted and observed
fluxes over $10^{16}$eV may be attributable to the fact that in a more
rigorous calculation  the leakage of  cosmic rays from  the galaxy
cannot be completely  neglected, while we assume here,  as a first
approximation, that  the residence time is  independent of energy.
In any case, the sharpness of the "knee" in the spectrum cannot be
explained  by smooth analytic  effects of  a transition  between a
regime dominated by diffusive propagation in the galactic magnetic
field and ones in which the  escape of cosmic rays from the galaxy
are suppressed (galactic modulation  models).  

In summary, we have
considered  the effect  of CR  particles interacting  with massive
neutrinos  in the  halo as  an explanation  of the  "knee"  in the
cosmic  ray spectrum.  We have  developed a  novel  approach which
allows  us to  obtain information  about the  properties  of relic
neutrinos in the galactic halo,  as well as about their masses and
magnetic dipole moments. We are able to reproduce the KASCADE data
around the  "knee" with  a mixed composition  of protons  and iron
nuclei,  with a  sharp  cutoff  in the  light  component which  is
compatible with experimental data.   Results of detailed fits will
be   presented  elsewhere,  together   with  discussions   of  the
possibilities  for earth-based  accelerator  experiments to  study
such neutrinos (or perhaps  other massive candidate particles). So
far,  the best fit  results are  not far  from the  present direct
limits\cite{pdg} set at  accelerators  ($5\times 10^{-7}\mu_B$), 
making  future prospects  very
interesting indeed!

\acknowledgements

We would like to thank our colleagues in the Pierre Auger
 Collaboration for many stimulating discussions, and in particular 
Lucas Taylor with whom early aspects of this idea were discussed. 
This work was supported by CONICET (Argentina) and the National 
Science Foundation (USA).

\newpage

\vskip 2 cm

\begin{center}
{\large Figure Caption}
\end{center}

\vskip 2 cm

Figure 1:  Differential cosmic ray spectrum as observed bt KASCADE. Solid line shows the result of the fit with a mixed proton-iron component.

\newpage

\begin{figure}
\label{plot}
\epsfig{file=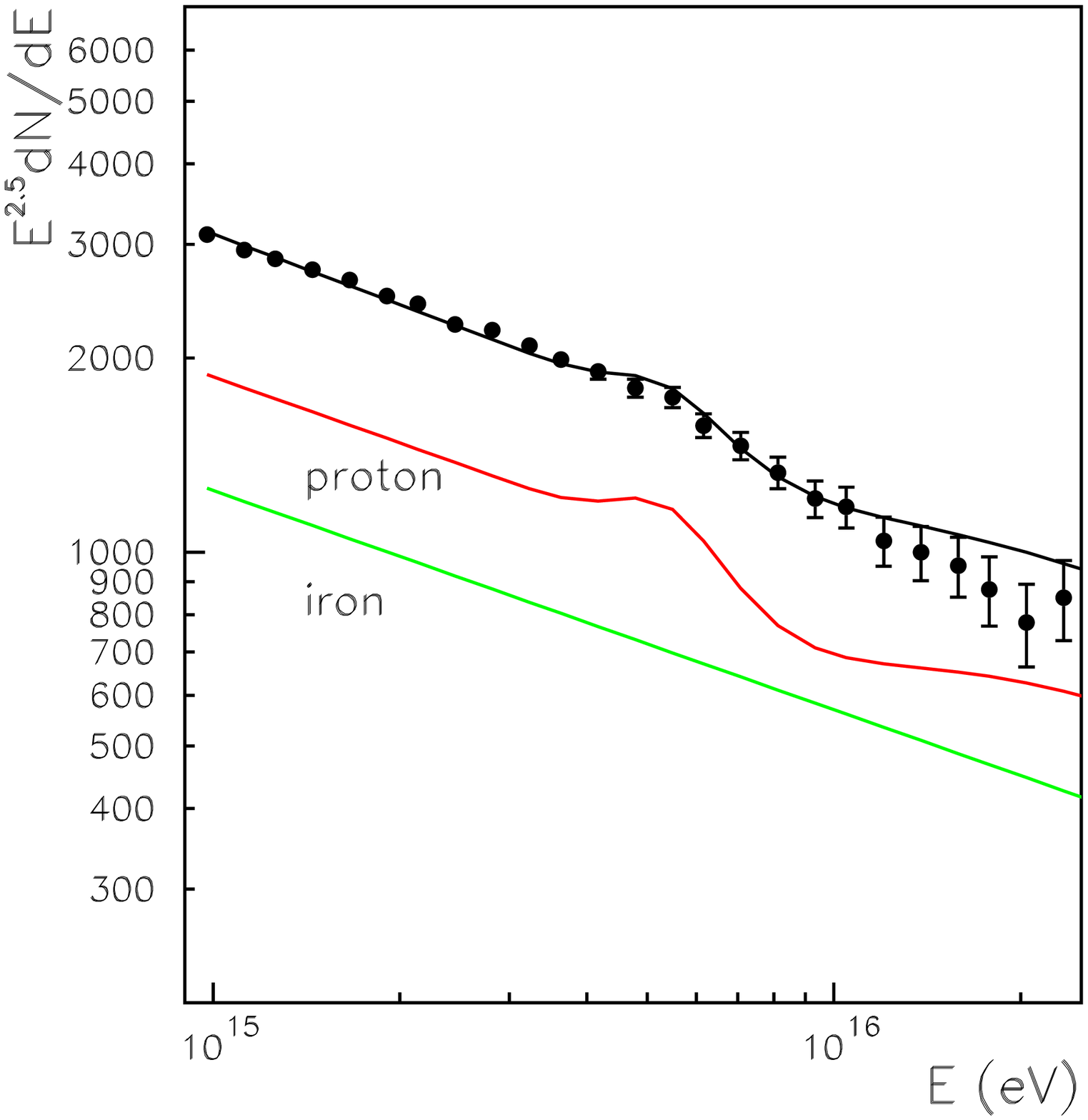,width=15cm,clip=}
\end{figure}

\end{document}